# Machine learning assisted speckle and OAM spectrum analysis for enhanced turbulence characterisation

WENJIE JIANG,[1,†] MINGJIAN CHENG,[1,†] LIXIN GUO,[1,*] XIANG YI,[2,] JIANGTING LI,[1] JUNLI WANG[1] AND ANDREW FORBES[1,3,*]

[1]School of Physics, Xidian University, South Taibai Road 2, Xi'an 710071 Shannxi, China
[2]School of Telecommunications Engineering, Xidian University, South Taibai Road 2, Xi'an 710071 Shannxi, China
[3]School of Physics, University of the Witwatersrand, Private Bag 3, Johannesburg 2050, South Africa
[†]These authors contributed equally.
*Lixin Guo (lxguo@xidian.edu.cn) or Andrew Forbes (andrew.forbes@wits.ac.za)

**Abstract:** Atmospheric turbulence degrades the performance of free-space optical (FSO) communication and remote sensing systems by introducing phase and intensity distortions. While a majority of research focuses on mitigating these effects to ensure robust signal transmission, an underexplored alternative is to leverage the transformation of structured light to characterize the turbulent medium itself. Here, we introduce a deep learning framework that fuses post-propagation intensity speckle patterns and orbital angular momentum (OAM) spectral data for atmospheric turbulence parameter inference. Our architecture, based on a modified InceptionNet backbone, is optimized to extract and integrate multi-scale features from these distinct optical modalities. This multimodal approach achieves validation accuracies exceeding 80%, substantially outperforming conventional single-modality baselines. The framework demonstrates high inference accuracy and enhanced training stability across a broad range of simulated turbulent conditions, quantified by varying Fried parameters ($r_0$) and Reynolds numbers (Re). This work presents a scalable and data-efficient method for turbulence characterization, offering a pathway toward robust environmental sensing and the optimization of dynamic FSO systems.

## 1. Introduction

Structured light, characterized by spatially tailored amplitude, phase, and polarization, is an established tool in applications ranging from high-capacity communications to advanced imaging and environmental sensing [1-3]. The capacity of these beams to encode and transport information through complex media makes them suitable for high-precision metrological tasks [4-6]. However, their deployment in atmospheric environments is impeded by stochastic turbulence-induced fluctuations that perturb the optical wavefront [7].

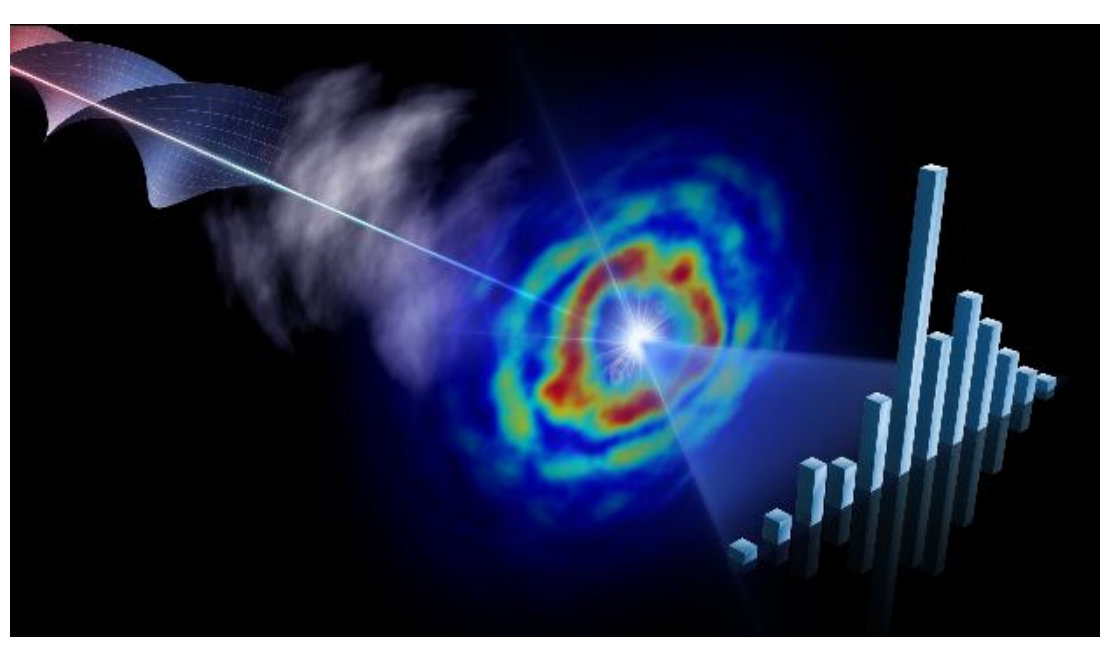

Fig.1 Schematic illustration of a vortex beam propagating through atmospheric turbulence. An incident beam with a well-defined OAM state experiences phase distortion and intensity fluctuation upon passing through a turbulent medium. This results in a speckled intensity pattern and a broadened OAM spectrum. These transformations provide distinct data modalities for inferring properties of the medium.

Atmospheric turbulence, arising from refractive index variations, induces degradative effects including beam wander, scintillation, and wavefront distortion, which collectively reduce signal fidelity [8]. As illustrated in Fig. 1, in free-space optical (FSO) systems that utilize vortex beams carrying orbital angular momentum (OAM), turbulence causes intermodal coupling, or crosstalk, which compromises transmission quality [9]. Consequently, accurate characterization of the turbulent channel is essential not only for optimizing system performance but also for using structured light as an active environmental probe. Foundational studies in this area often simplify the problem by assuming static or quasi-static turbulence, allowing for the isolated investigation of key atmospheric parameters.

Recent work on “turbulence eigenmodes”, structured fields resilient to specific turbulence statistics, presumes prior knowledge of the turbulence profile, which is rarely available in dynamic settings [10]. This limitation highlights the need for data-driven turbulence characterization frameworks. Vortex beams, with their helical phase fronts and quantized OAM states, are highly sensitive to phase perturbations, making them effective candidates for diagnostic sensing [11-12]. The turbulence-induced distortions encoded in their post-propagation intensity and phase can be leveraged to infer the state of the medium. Experimental studies have demonstrated the feasibility of using vortex beams for environmental monitoring, where metrics like ring deformation and centroid displacement correlate with turbulence strength ($C_n^2$) [13]. Furthermore, techniques such as the rotational Doppler effect have been used for flow velocimetry [14-15], while other beam engineering methods have improved refractive index measurements [16]. These developments provide a foundation for data-driven approaches, where machine learning can be used to interpret the complex patterns arising from these light-matter interactions [17].

Deep learning, particularly the use of convolutional neural networks (CNNs), has been successfully applied to turbulence characterization by analyzing distorted speckle patterns [18-21]. Such models can accurately classify turbulence levels or perform regression on the $C_n^2$ value by treating intensity images as information-rich feature maps [22-24]. In parallel, the OAM spectrum has been used as a diagnostic tool, with models trained on OAM spectral data achieving high accuracy in estimating environmental parameters [25-26]. However, most existing machine learning approaches are unimodal, relying on either intensity patterns or OAM spectra. This dependence on a single modality can limit the diversity of extracted features and increase the risk of overfitting, particularly in data-constrained scenarios [27]. In contrast, a multimodal strategy that fuses spatial (intensity) and modal (OAM) features can exploit complementary information. Intensity speckles capture the cumulative spatial effects of turbulence, while the OAM spectrum reveals details about the induced phase dynamics [4]. Multimodal fusion, therefore, presents an opportunity to improve model generalizability and performance.

Here, we propose a dual-input deep learning framework that concurrently processes the post-propagation intensity speckle patterns and OAM spectra of vortex beams to infer atmospheric turbulence parameters. Specifically, we classify turbulence regimes defined by the Fried parameter ($r_0$) and Reynolds number (Re) under simulated static conditions. We investigate two fusion architectures: a single-output model using feature concatenation and a dual-output, multi-task learning model. We benchmark these architectures against a standard intensity-only CNN and demonstrate that our multimodal fusion designs, particularly the dual-output model, yield substantial improvements in classification accuracy, training stability, and robustness. This work provides a foundation for developing real-time characterization methods for dynamic atmospheric turbulence.

## 2. Vortex Beam Propagation Modeling Through Atmospheric Turbulence

To construct a robust dataset for validating the proposed deep learning framework, we performed numerical simulations of vortex beam propagation through atmospheric turbulence.

The optical field was modeled as a Laguerre-Gaussian (LG) beam with a topological charge of $l$=3, a widely used vortex beam mode due to its well-defined OAM characteristics. Simulations assumed static or quasi-static turbulence conditions over the typical data acquisition period. This assumption enabled isolated characterization of the system's sensitivity to stable or slowly varying turbulence parameters, such as the Fried parameter and Reynolds number, without the complexities introduced by rapid temporal fluctuations.

Atmospheric turbulence is modeled as a random field of refractive index fluctuations. The statistical properties of these fluctuations are well-approximated by the Kolmogorov theory, parameterized by metrics quantifying turbulence strength and scale. The Fried parameter ($r_0$), a key metric of wavefront coherence, quantifies the effective diameter over which an optical wavefront remains largely undistorted. It was calculated from the refractive index structure parameter ($C_n^2$) using the integral expression:

$$r_0 = \left(0.423k^2 \int_0^L C_n^2(z)\,dz\right)^{-3/5} \quad (1)$$

where $k=2\pi/\lambda$ is the wave number, and $L$ is the propagation path length. This formulation reflects the spatially accumulated wavefront aberration effects induced by inhomogeneous turbulence along the beam path.

Turbulence conditions were further characterized by the Reynolds number (Re), a dimensionless quantity related to the relative magnitudes of inertial and viscous forces in a fluid flow. In the context of atmospheric turbulence, it can be defined in relation to the outer and inner scales of turbulence, $L_0$ and $l_0$, respectively. A common relation is $\mathrm{Re}=(L_0/l_0)^{4/3}$. As Re increases, the energy cascade spans a broader range of eddy sizes, leading to greater phase fluctuations and enhanced beam distortion. This wide spectral distribution of turbulence across scales influences both the intensity speckle and OAM spectrum, providing distinct spatial and modal features for learning-based turbulence inference.

**Table 1. Atmospheric Turbulence Strength (Scintillation Index, $\sigma_I^2$) for Selected Fried Parameters ($r_0$) and Reynolds Numbers (Re).**

| Scintillation index $\sigma_I^2$ | | $D/r_0$ | | | |
|---|---|---|---|---|---|
| | | 40 | 20 | 13 | 10 |
| Re | 1574 | 5.062 | 1.594 | 0.811 | 0.502 |
| | 2000 | 6.166 | 1.942 | 0.988 | 0.612 |
| | 3960 | 8.964 | 2.824 | 1.437 | 0.889 |
| | 10000 | 11.315 | 3.564 | 1.813 | 1.123 |
| | 20000 | 12.186 | 3.839 | 1.953 | 1.209 |

The strength of atmospheric turbulence in our simulations was additionally characterized using the scintillation index ($\sigma_I^2$), which quantifies the normalized variance of intensity fluctuations. The values of $\sigma_I^2$ presented in Table 1 were calculated using the following analytical expression, applicable across weak to strong turbulence regimes:

$$\sigma_I^2 = 3.86\sigma_R^2 \left[ \left(1+1/Q_m^2\right)^{\frac{11}{12}} \sin\left(\frac{11}{6}\tan^{-1} Q_m\right) - \frac{11}{6} Q_m^{-5/6} \right] \quad (2)$$

Here, $\sigma_R^2$ is the Rytov variance, given by $\sigma_R^2 = 1.23 C_n^2 k^{7/6} L^{11/6}$. The parameter $Q_m$ is defined as $Q_m = 35.04L/\left(kl_0^2\right)$. For each turbulence condition, $C_n^2$ was derived from the specified $r_0$, and $l_0$ was implicitly determined by the Reynolds number given a fixed outer scale $L_0$.

To cover a representative range of turbulence conditions, we modeled 20 distinct turbulence levels by varying the Fried parameter ($r_0$) from 5 mm (strong turbulence) to 20 mm (moderate-to-weak turbulence), corresponding to Reynolds numbers (Re) in the range of 1574 to 20000, under a fixed outer scale $L_0$ = 10 m. The resulting $\sigma_I^2$ values, as shown in Table 1, indicate conditions ranging from weak ( $\sigma_I^2$ <1), through moderate ( $\sigma_I^2$ ≈1), to strong ( $\sigma_I^2$ >1) turbulence. While the scintillation index offers a valuable metric for turbulence strength, the Fried parameter ($r_0$) and Reynolds number (Re) remain the primary ground truth labels for training and evaluating our machine learning model, as these directly quantify the wavefront coherence and the scale distribution of the turbulence that our model aims to infer through classification.

Beam propagation was simulated using a multi-phase-screen model, as depicted in Fig. 2. This model emulates the cumulative impact of atmospheric turbulence by introducing a series of statistically independent thin phase screens along the propagation path. Each phase screen represents localized refractive index fluctuations, enabling the model to capture both small-scale and large-scale turbulent effects. This approach aligns with the Kolmogorov theory of turbulence, providing a physically sound approximation of random inhomogeneities in the atmospheric refractive index field.

To model beam evolution, we employed the split-step Fourier method, which alternates between free-space propagation in the spectral (Fourier) domain and phase modulation due to turbulence in the spatial domain. This method balances numerical efficiency and physical accuracy for long-range beam propagation modeling [28-29]. The turbulence-induced phase distortions were generated using the Von Kármán turbulence power spectral density, which incorporates both outer and inner scale effects. The spectral representation was implemented via the Fast Fourier Transform (FFT), allowing the phase screens to reflect spatial frequency content consistent with realistic turbulence. For implementation details, refer to [28].

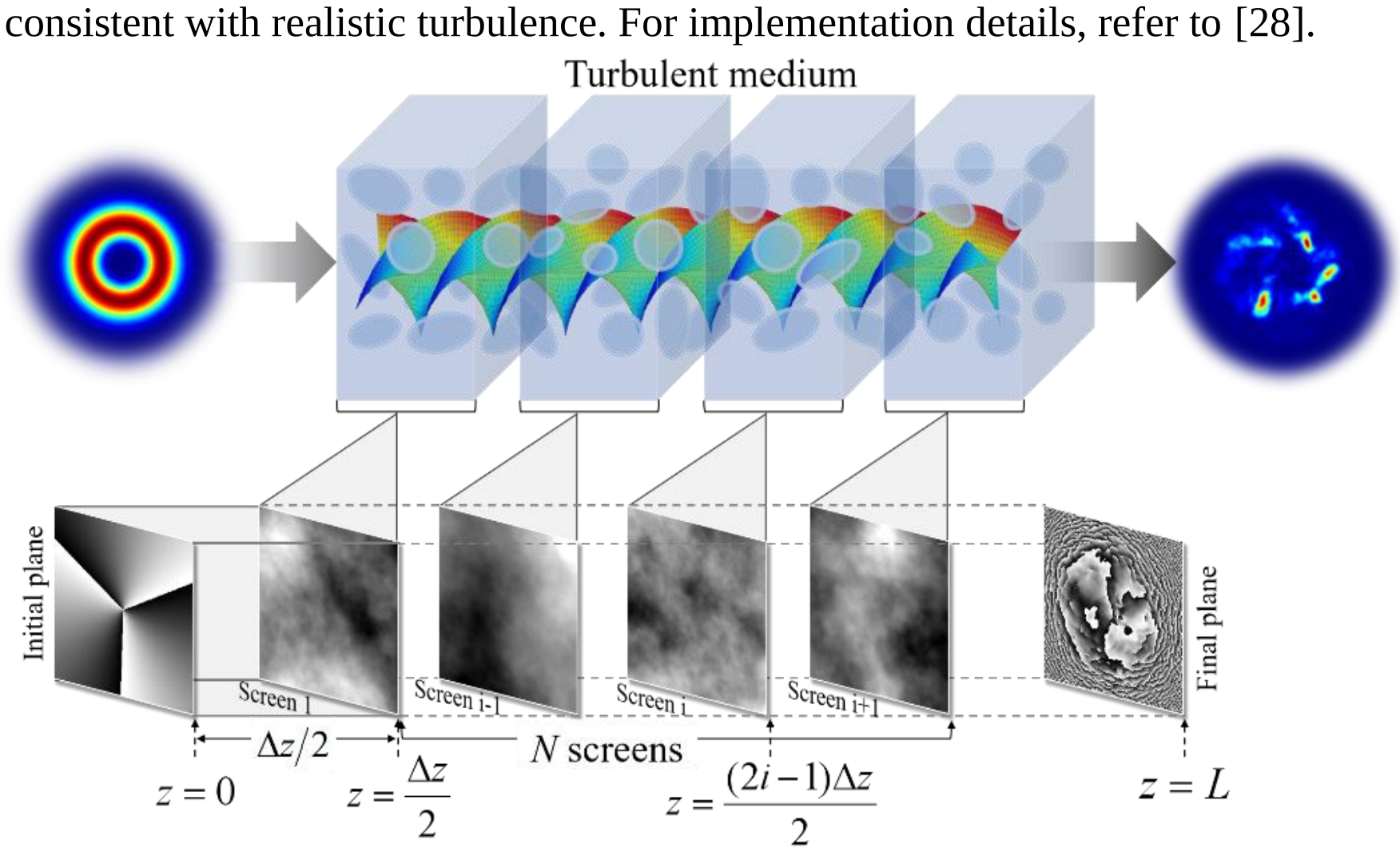

Fig. 2. Schematic diagram of the multi-phase screen model for simulating vortex beam propagation through atmospheric turbulence.

**Table 2. Simulation parameters for vortex beam propagation under atmospheric turbulence.**

| Parameter | Value | Parameter | Value |
|---|---|---|---|
| Path Length | $L$ = 1500 m | Outer Scale | $L_0$ = 10 m |
| Phase Screen Size | 512×512 pixels | Aperture Diameter | $D$ = 0.2 m |
| Phase Screen Number | $N$ = 150 | Beam radius | $w_0$ = 0.02 m |
| Wavelength | $\lambda$ = 633.2 nm | OAM mode | $l$ = 3 |

Simulation parameters are summarized in Table 2. The total propagation distance was $L$ = 1500 m, with 150 phase screens uniformly distributed along the path to ensure sufficient spatial resolution of the turbulence dynamics. Each phase screen had a resolution of 512×512 pixels, covering an aperture of 0.2 m. The beam waist radius was set to $w_0$ = 0.02 m, and the operating wavelength was $\lambda$ = 633.2 nm. For each simulated turbulence level, 150 statistically independent phase screen realizations were generated. From the complex optical field obtained after structured light propagation through these turbulent layers, two distinct data modalities were derived.

First, far-field intensity speckle patterns were generated by numerically propagating the complex optical field into the Fraunhofer regime. The resulting two-dimensional intensity distributions were recorded as images. For computational efficiency, these images (419×513 pixels, RGB) were converted to grayscale and resized to 150×150 pixels. This resolution was selected to balance feature preservation with computational demands.

Second, the OAM spectrum was extracted from the same complex optical field. This process quantifies the redistribution of the initial OAM state of the structured light beam due to turbulence-induced phase aberrations. The method, often termed spiral spectrum expansion, involves decomposing the complex optical field into its fundamental orthogonal OAM constituent modes. Each OAM mode, uniquely defined by an integer topological charge and a helical phase front ($\exp(im\phi)$), was isolated by projecting the complex field onto that mode's basis. This projection quantifies the power contained within each mode. Measurements covered topological charges from -20 to +19 (40 discrete OAM modes), a range chosen to capture the typical OAM modal spread under varying turbulence strengths. These power values formed the OAM spectrum.

Both the preprocessed grayscale images and the extracted OAM spectra were normalized to the range [0, 1] using min-max scaling. This standardization ensures consistent data scaling across modalities, which can facilitate neural network training. The complete dataset comprised 4800 samples, uniformly distributed across 20 turbulence classes (approximately 240 samples per class). To ensure a balanced distribution across subsets and minimize potential bias, the dataset was randomly partitioned into training, validation, and test subsets using a 4:1:1 ratio. This training dataset size, with 160 samples per class, is representative of data-constrained scenarios encountered in some real-world deep learning applications. Representative samples illustrating these optical responses across varying turbulence strengths are presented in Figure 3.

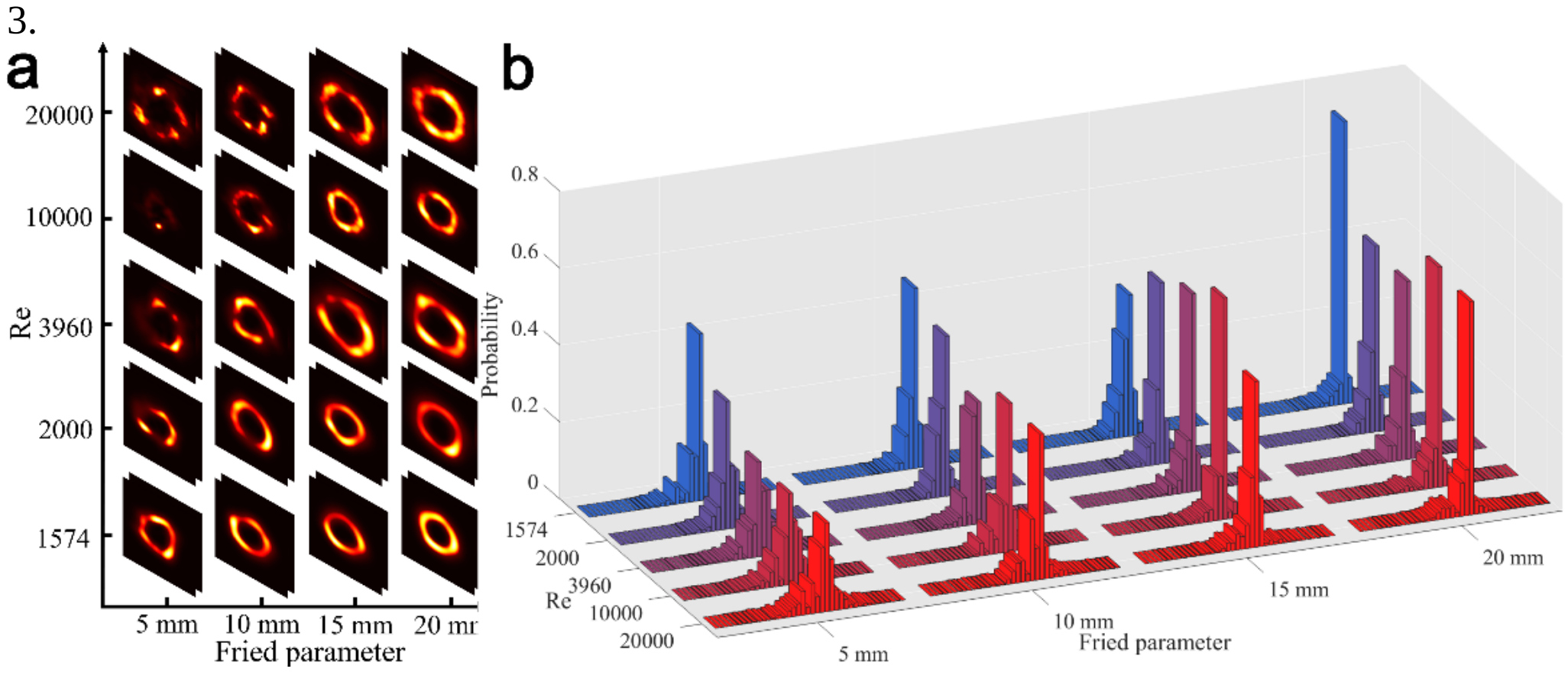

Fig. 3. Turbulence-induced effects on vortex beam characteristics. (a) Far-field intensity distributions showing characteristic speckle patterns, and (b) corresponding OAM spectral broadening, demonstrating turbulence-induced mode coupling. These examples illustrate the observed changes in speckle formation and OAM mode coupling across varying turbulence strengths.

## 3. Turbulence Detection Method Based on Convolutional Neural Networks (CNN)

This section details the architectural design and implementation of three CNN models developed for atmospheric turbulence classification. We first describe the foundational network backbone, followed by the specific implementations of a baseline intensity-only model and two multimodal fusion architectures that integrate OAM spectral data.

Atmospheric turbulence introduces complex, multi-scale distortions to a propagating optical beam, resulting in features that range from large-scale structural deformation to fine-grained speckle. To effectively classify these patterns, a model must extract hierarchical features across these different spatial scales.

For this task, the Inception architecture was selected as the foundational backbone for all models [30]. The core principle of Inception is its use of multi-branch modules that apply convolutional filters with varying kernel sizes (e.g., 1×1, 3×3, 5×5) in parallel within a single layer. This structure enables the network to simultaneously capture fine-grained textural information and broader spatial patterns, making it well-suited for analyzing turbulence-induced optical fields. Furthermore, the strategic use of 1×1 convolutions for dimensionality reduction within each module significantly reduces the computational cost and parameter count, rendering the architecture efficient for processing high-dimensional optical data.

The baseline model (Model 1), depicted in Figure 4(a), was developed to establish a performance benchmark using only intensity speckle images. Its architecture is a lightweight framework inspired by InceptionNet. A preprocessed grayscale input image first passes through a 1×1 convolutional layer with 16 filters to increase channel depth. The resulting feature maps are then processed by a stack of four Inception modules, with two max-pooling layers interleaved to downsample the spatial dimensions, extract increasingly abstract hierarchical features, and enhance feature selectivity to avoid the dilution of important features by the next global average pooling (GAP) layer. The output of this convolutional backbone is a 19×19×128 feature map. A GAP layer then reduces this map to a 128-dimensional feature vector. This vector is processed by a classification head composed of two fully connected (FC) layers with ReLU activation and a final FC layer with a Softmax function to yield a probability distribution over the 20 turbulence classes.

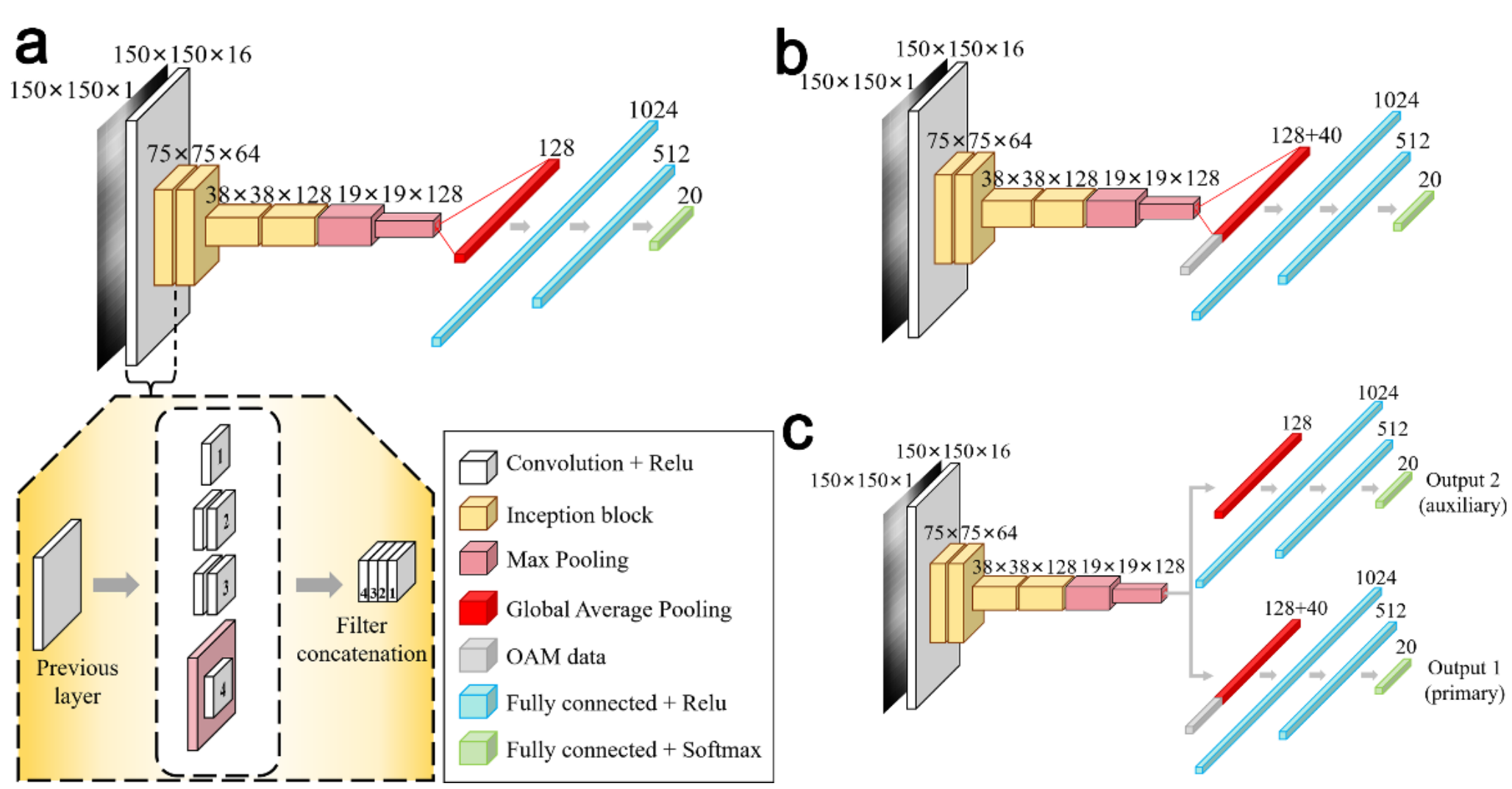

Fig. 4. Schematics of the three proposed CNN architectures. (a) Model 1 (Baseline): A single-input CNN processing only intensity speckle images. (b) Model 2 (Single-Output Fusion): A dual-input model that concatenates intensity-derived features with OAM spectral data before final classification. (c) Model 3 (Dual-Output Fusion): A dual-input model employing a multi-task learning framework with a primary multimodal output and an auxiliary intensity-only output to regularize the shared feature extractor.

To leverage the complementary information contained within both the spatial intensity distribution (amplitude) and the OAM spectrum (phase), we developed two fusion models. The single-output fusion model (Model 2), shown in Figure 4(b), integrates OAM spectral data at the feature level. It utilizes the same feature extraction backbone as Model 1 to generate a 128-dimensional image feature vector. This vector is then concatenated with the 40-dimensional normalized OAM spectrum vector, forming a fused 168-dimensional multimodal representation. This fused vector is subsequently passed to a classification head, identical to that in Model 1, to predict the turbulence class. For comparison, a shallow multi-layer perceptron (MLP) was trained exclusively on OAM spectrum data. The resulting low classification accuracy, as shown in Figure 5, indicates that OAM spectral data alone are insufficient for robust turbulence characterization, thereby motivating the fusion with spatial features from the intensity speckle patterns.

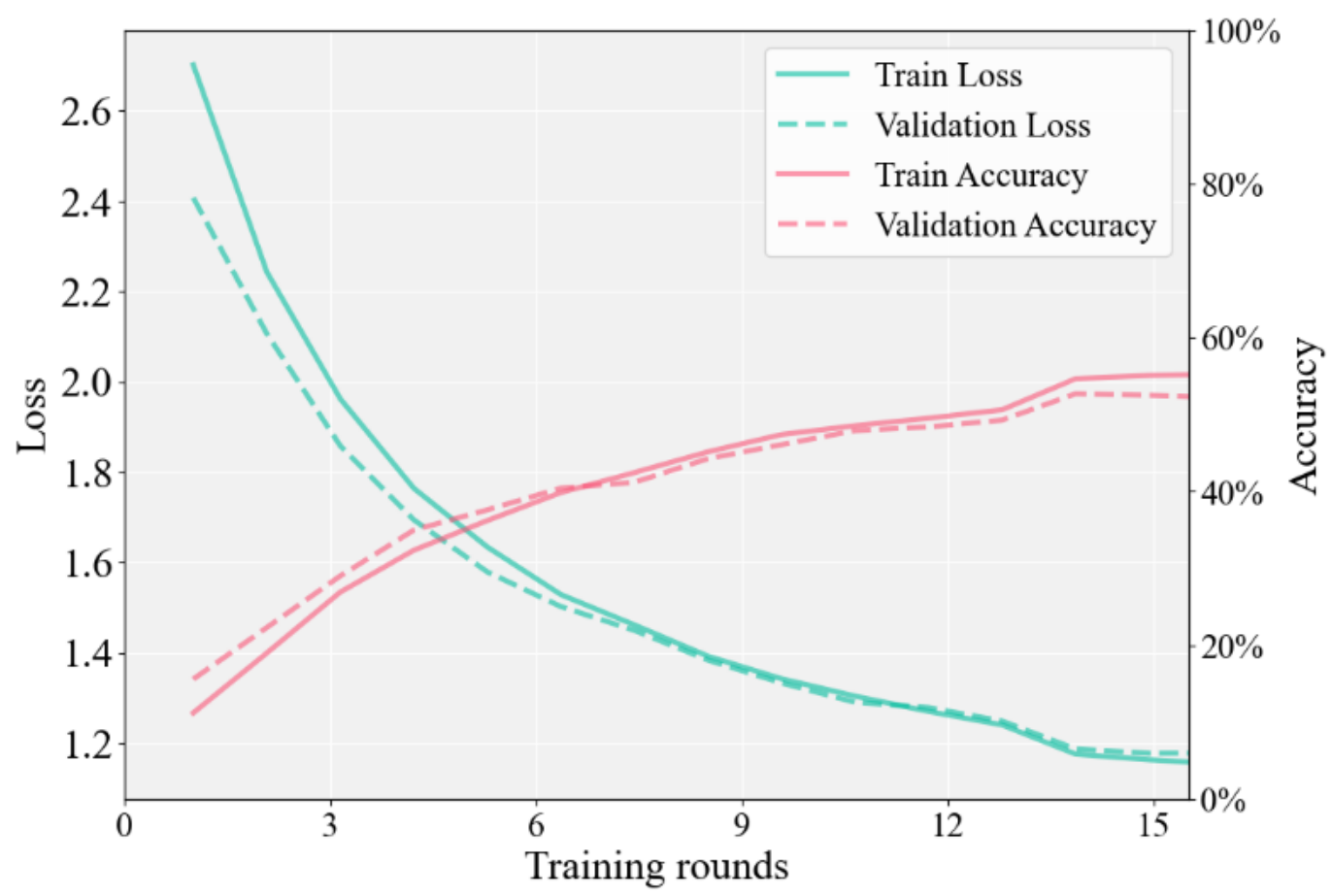

Fig. 5. Training and validation curves for a network trained solely on OAM spectrum data. The model's low final accuracy underscores the value of the spatial information provided by intensity images.

The dual-output fusion model (Model 3), depicted in Figure 4(c), implements a multi-task learning framework aimed at enhancing feature representation robustness and improving generalization capabilities. The architecture comprises a shared convolutional backbone, identical to that of Model 1, which branches into two distinct classification heads. The primary output branch is responsible for the main multimodal classification task. It receives a 168-dimensional fused feature vector, formed by concatenating the 128-dimensional image features from the shared backbone with the 40-dimensional normalized OAM spectrum vector. The classification head within this branch is architecturally identical to that employed in Model 1. The output of this branch yields the final probability distribution over turbulence classes during inference. The auxiliary output branch serves as a regularization mechanism for the shared feature extractor. It processes only the 128-dimensional image feature vector, performing an auxiliary classification task based solely on intensity information, with its classification head also identical to that of Model 1.

During training, the auxiliary branch imposes an inductive bias on the shared backbone. By requiring the shared features to be discriminative for the unimodal (intensity-only) classification task, it effectively mitigates the risk of the network overfitting to idiosyncratic features of the fused multimodal input. This regularization encourages the learning of more robust and generalizable feature representations within the shared layers. The total objective function for Model 3 is defined as a weighted sum of the primary ($L_{\text{main}}$) and auxiliary ($L_{\text{aux}}$) classification losses: $L_{\text{total}}=w_1L_{\text{main}} +w_2L_{\text{aux}}$, where $w_1$ and $w_2$ are hyperparameters that control the relative contribution of each task to the overall loss. An empirical evaluation of various weight combinations was conducted to determine suitable values for $w_1$ and $w_2$. As illustrated in Figure 6, a configuration of $w_1$=0.7 and $w_2$=0.4 was empirically found to yield a favorable balance between training convergence rate and final classification accuracy.

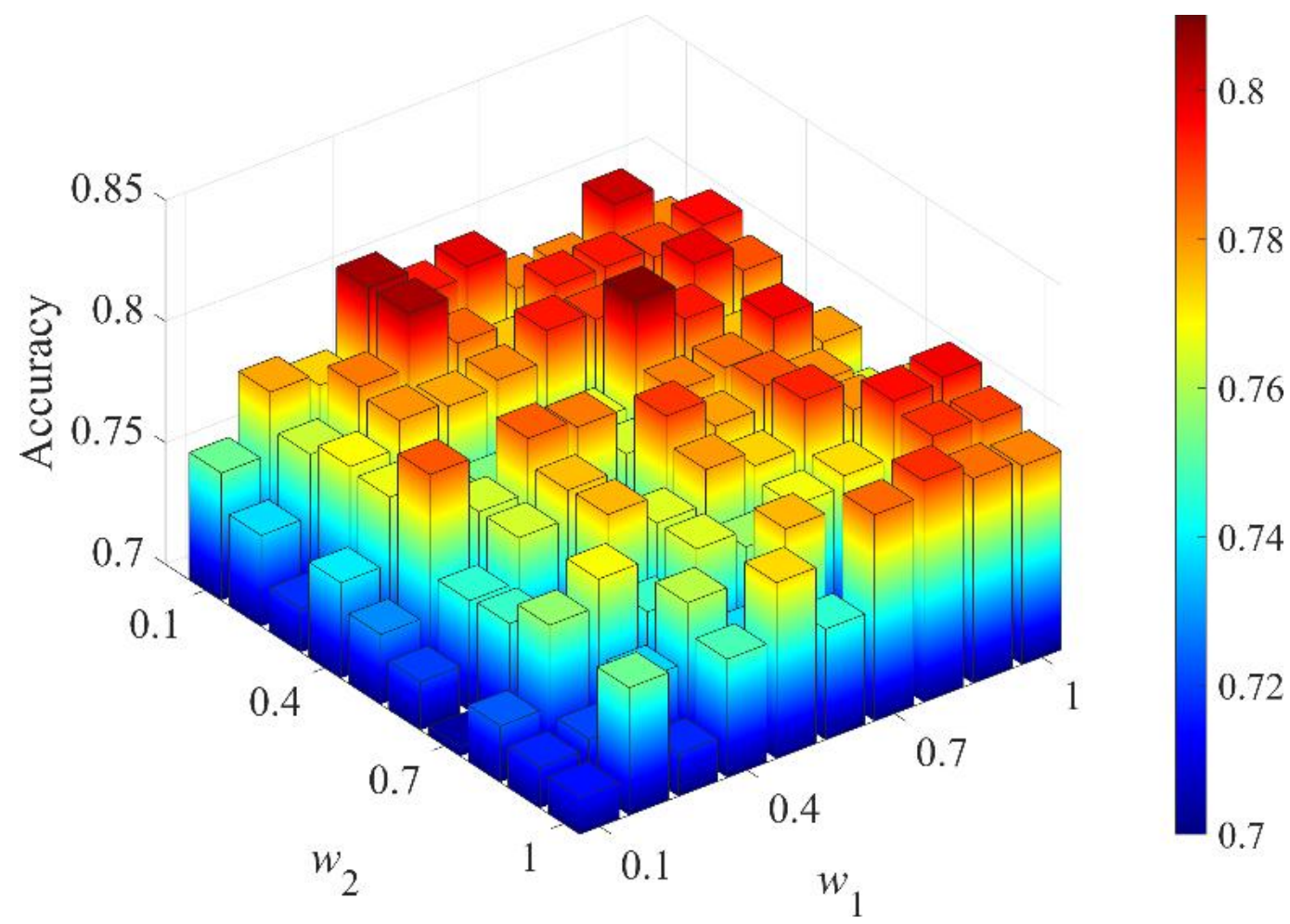

Fig. 6. Median classification accuracy as a function of loss weights $w_1$ and $w_2$. Data points represent the median accuracy over 30 independent runs. The configuration ($w_1$=0.7 and $w_2$=0.4) yielded the highest stable performance.

All models were implemented in Python using the TensorFlow and Keras libraries. The models were trained using the Adam optimizer with an initial learning rate of $10^{-3}$. A step decay learning rate schedule was applied, reducing the rate by a factor of 0.3 every 14 epochs to facilitate fine-tuning during later training stages. The sparse categorical cross-entropy was used as the loss function, defined as:

$$\text{Loss} = -\frac{1}{N}\sum_{i=1}^{N}\log\left(p_i, y_i\right) \tag{3}$$

where $N$ is the number of classes, $y_i$ is a binary indicator (1 for the correct class, 0 otherwise), and $p_i$ is the model's predicted probability for class $i$.

Each model was trained for 15 epochs with a batch size of 32. It was observed that the models achieved stable validation accuracy without requiring explicit regularization techniques such as Dropout. This suggests the model capacity was well-matched to the complexity of the dataset used. All computations were performed on a workstation equipped with an Intel Core i9-13900K CPU and an NVIDIA GeForce RTX 3060 GPU.

## 4. Results and Analysis

The performance of the proposed OAM spectrum-assisted fusion strategy was evaluated in Fig. 4 through comparative experiments on three distinct network architectures: (1) Model 1, a baseline CNN using only intensity speckle images; (2) Model 2, a dual-input, single-output fusion model combining intensity images and OAM spectrum data; and (3) Model 3, a dual-input, dual-output architecture incorporating an auxiliary classification objective designed to enhance feature representation. To ensure statistical validity, each model was independently trained 30 times on the shuffled dataset for cross-validation comparison.

After 15 training epochs, the baseline (Model 1) achieved a mean validation accuracy of 67.41% with an associated loss of 0.8022. The integration of OAM spectral data in Model 2 resulted in a validation accuracy of 78.11% (a 10.7% increase), and a reduced loss of 0.5289. Model 3 yielded the highest performance, with a mean validation accuracy of 80.68% and the lowest mean loss of 0.4830. These results (summarized in Fig. 7a-b) indicate that the multimodal fusion approach is effective for turbulence classification, particularly in data-constrained scenarios.

The computational overhead associated with the fusion models was minimal. The average training times per session were comparable across architectures: 94.3 s (Model 1), 95.4 s (Model 2), and 96.5 s (Model 3). This demonstrates that the observed performance gains were achieved with negligible additional computational cost.

Model stability was assessed by analyzing the variance in validation accuracy across the 30 independent training runs. The standard deviation of validation accuracy decreased progressively from 0.071 for Model 1 to 0.042 for Model 2, and to 0.020 for Model 3. This corresponds to a variance reduction of over an order of magnitude from Model 1 ($4.97\times10^{-3}$), Model 2 (variance = $1.73\times10^{-3}$), to Model 3 ($3.96\times10^{-4}$), indicating that the dual-output fusion architecture confers substantially enhanced training stability.

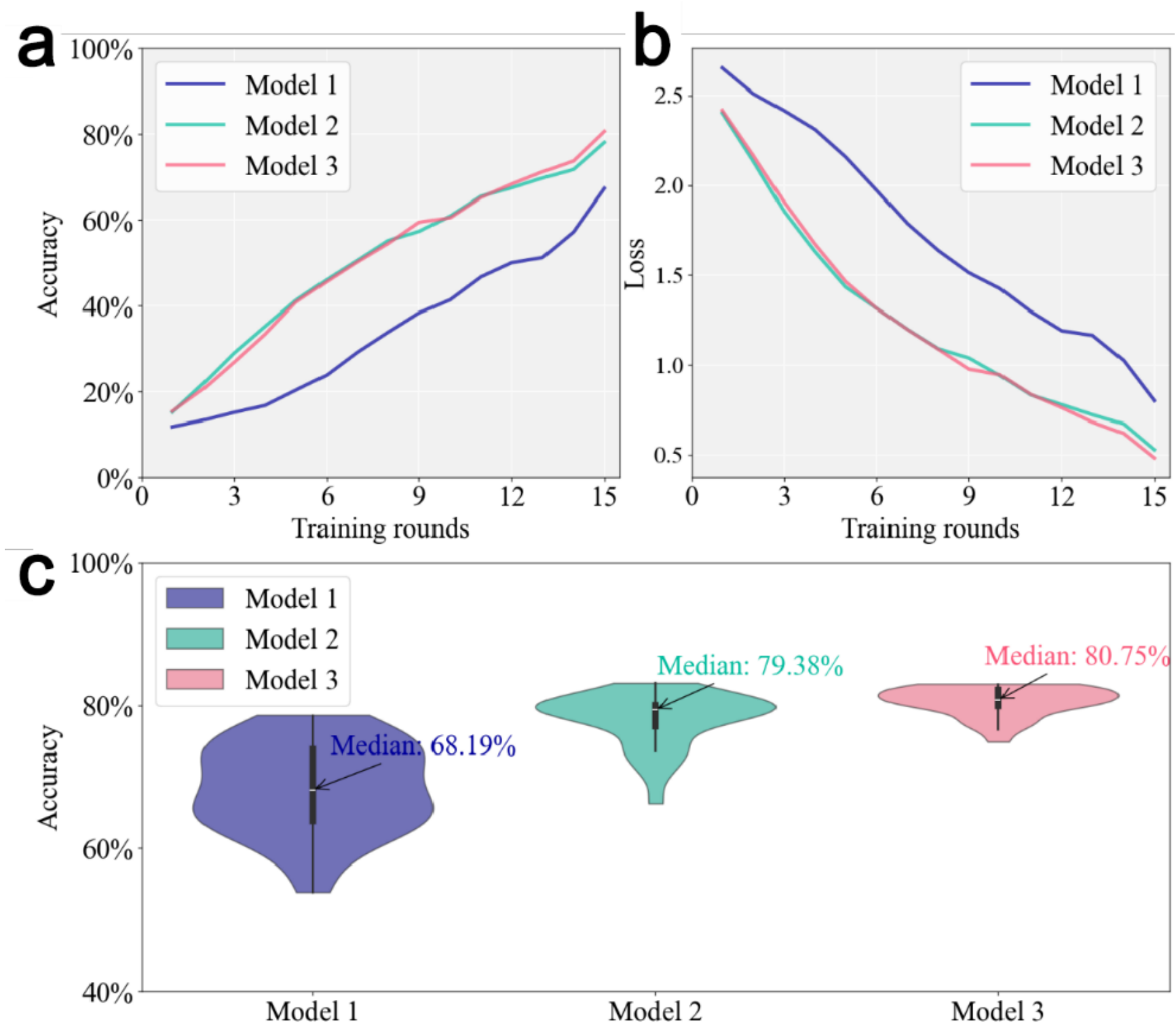

Fig. 7. Model performance evaluation across 30 independent training runs. (a) Mean validation accuracy and (b) mean validation loss. (c) Boxplot distribution of test accuracies for all trained models. Model 3, the dual-input, dual-output architecture, demonstrates the highest accuracy and lowest variance.

To evaluate generalization, all 90 trained models (30 for each architecture) were evaluated on an unseen test dataset. The distributions of test accuracies are presented in the boxplot in Figure 7(c). The median test accuracies were 68.19% (Model 1), 79.38% (Model 2), and 80.75% (Model 3). Model 3 exhibited the narrowest interquartile range and overall spread, which suggests superior generalization and stability when applied to new data.

To analyze model performance on a class-specific basis, normalized confusion matrices were generated for representative models (Fig. 8a-c). The matrices illustrate classification performance across turbulence conditions defined by the Reynolds number (Re) and the ratio of the aperture diameter to the Fried parameter ($D/r_0$). Both fusion models (2 and 3) show increased diagonal dominance compared to the baseline, signifying improved classification accuracy across all turbulence regimes.

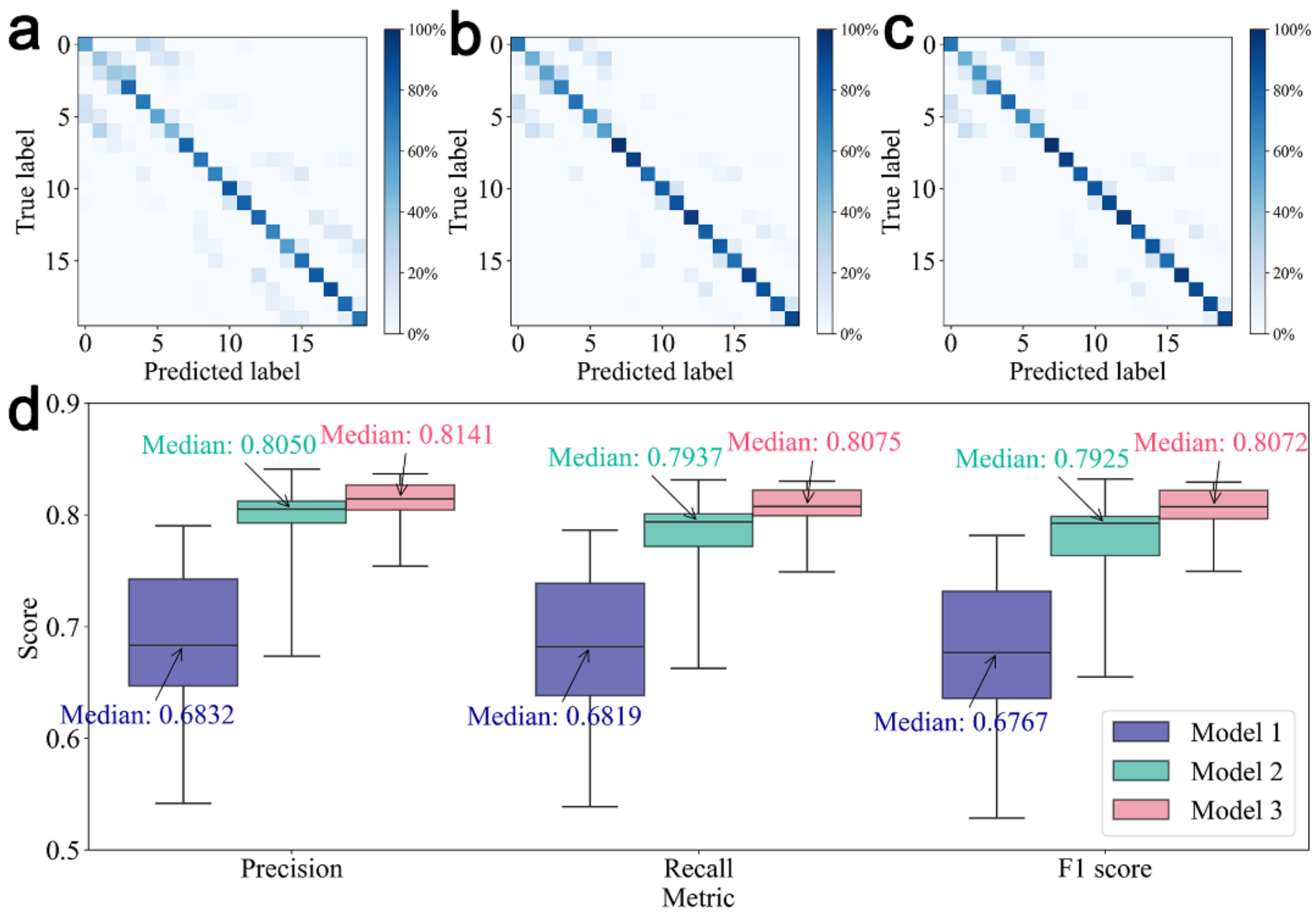

Fig. 8: Normalized confusion matrices for (a) Model 1, (b) Model 2, and (c) Model 3. (d) Boxplot distributions of Precision, Recall, and F1 Score across 30 test runs for all models.

This qualitative observation was quantified using Precision, Recall, and F1 Score, aggregated over all 30 test runs (Fig. 8d). Model 3 consistently outperformed Models 1 and 2 across all three metrics. This result confirms the benefit of fusing spatial-domain intensity features with phase-derived OAM spectral data and suggests that the auxiliary learning objective in Model 3 effectively improves the discriminability of the learned features.

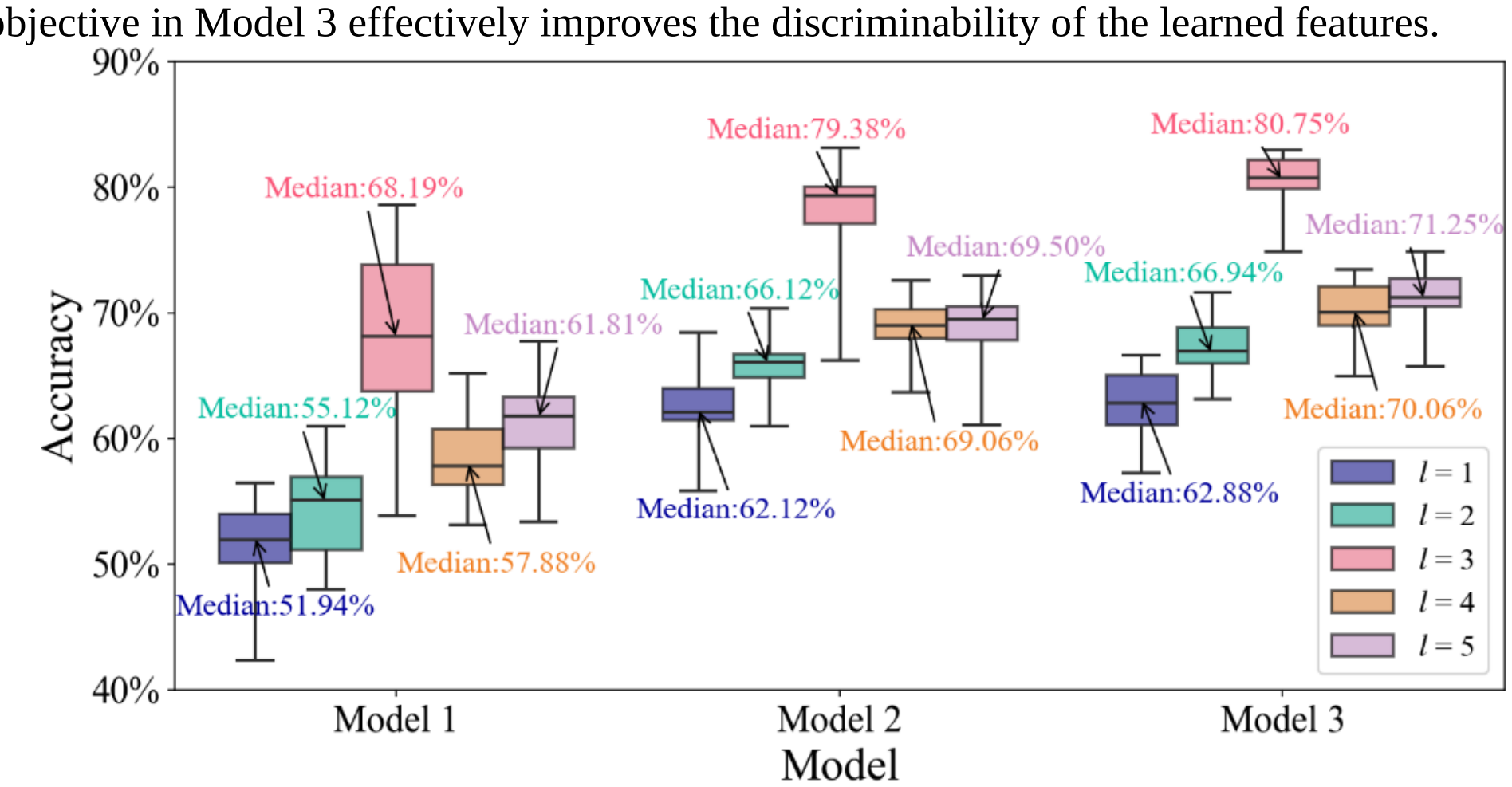

Fig. 9. Boxplot distributions of test accuracy as a function of incident OAM topological charge ($l$). All models exhibit peak performance at $l$=3.

The influence of the incident OAM topological charge, $l$, on classification accuracy was investigated by evaluating all models with $l$ ranging from 1 to 5. As shown in Figure 9, the fusion models consistently yielded higher classification accuracies than the baseline across all tested values of $l$. A key observation is that all three model architectures achieved their optimal performance at $l$=3. The consistent peak performance at this specific topological charge suggests that $l$=3 may represent an optimal parameter for probing turbulence characteristics within the parameter space investigated in this study.

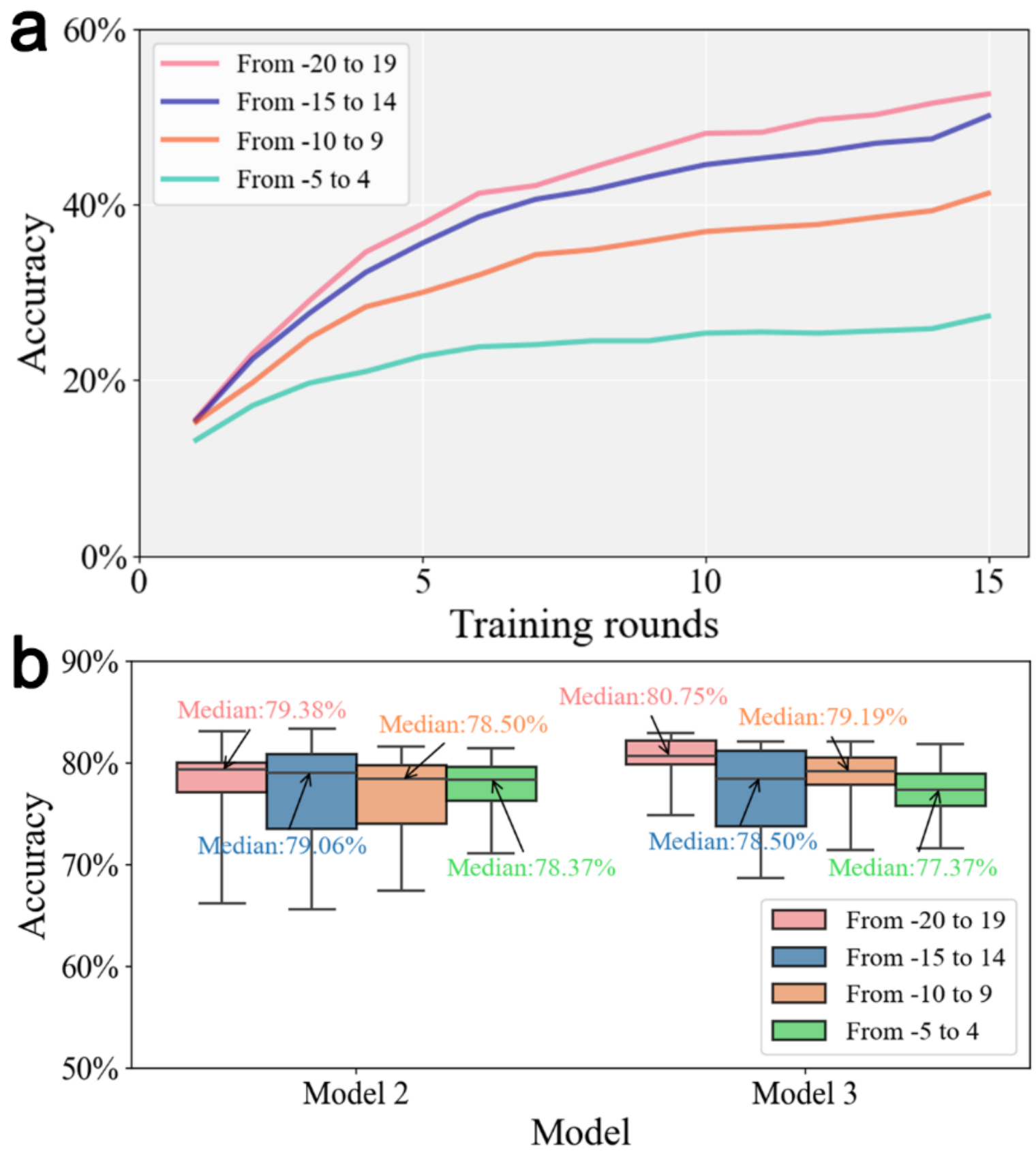

Fig. 10. Impact of OAM spectral range on performance. (a) Accuracy of an OAM-only network declines as the spectral range is truncated. (b) Test accuracy of fusion models with varying OAM spectral ranges. The performance of Model 3, unlike Model 2, is sensitive to the information content of the OAM spectrum.

Subsequently, to dissect the contribution of the OAM data, the effect of its spectral range on classification performance was analyzed in Figure 10. The information content of the OAM spectrum was first benchmarked using a standalone fully connected network. As shown in Figure 10(a), reducing the spectral range from 40 modes ($l \in [-20, +19]$) down to 10 modes ($l \in [-5, +4]$) resulted in a monotonic decrease in classification accuracy, confirming that a wider spectrum contains more discriminative information.

Next, the impact of this truncation was evaluated in Fig. 10(b) on the fusion models. The performance of Model 2 remained largely unaffected by the reduction in OAM spectral range, suggesting its performance is dominated by features extracted from the speckle intensity images. In contrast, Model 3's accuracy exhibited a clear degradation when the spectrum was limited to 10 modes. This is consistent with its architecture, where a dedicated branch processes the OAM spectrum. Reducing the information content of that input directly attenuates the efficacy of the branch, thereby diminishing the performance benefit derived from the multimodal fusion. Consequently, a spectral range of $l \in [-20, +19]$ was adopted for all primary experiments to maximize the information available to the fusion models.

## 5. Conclusion

This study presents a deep learning framework for the classification and preliminary regime identification of atmospheric turbulence, leveraging the unique transmission characteristics of vortex beams. The proposed approach addresses the inherent challenge of limited data availability by employing a dual-branch CNN architecture. This architecture effectively fuses post-propagation intensity speckle patterns with OAM spectral data, thereby mitigating a significant practical constraint. The integration of OAM spectral features yielded enhanced classification performance, with the dual-output fusion model achieving an average validation accuracy of approximately 81% for turbulence regime identification. Furthermore, the developed framework exhibited notable training stability and generalization capabilities, performing effectively even under data-constrained training conditions and with limited epochs.

This work establishes a deep learning method for rapid classification and initial regime identification of atmospheric turbulence, specifically inferring the Fried parameter ($r_0$) and Reynolds number (Re) under simulated static or quasi-static conditions. These findings underscore the potential of deep learning, combined with structured light, for advancing atmospheric turbulence characterization.

Future research directions should focus on extending this methodology to include validation against real or simulated laboratory-based turbulence experiments and to address the characterization of dynamically evolving turbulence. This will necessitate the development of capabilities for precise and continuous parameter retrieval via regression models, enhancement of temporal measurement resolution, implementation of real-time data processing, exploration of adaptive sampling strategies, and investigation of additional physical dimensions (e.g., phase, polarization). Such advancements are critical for improving characterization accuracy and enabling the practical applicability of this framework in dynamic free-space optical systems.

**Funding.** 111 Project (B17035); National Natural Science Foundation of China (U20B2059, 62231021, 61621005, 62201613); Shanghai Aerospace Science and Technology Innovation Foundation (SAST-2022-069); Fundamental Research Funds for the Central Universities (ZYTS25121).

**Disclosures.** The authors declare no conflicts of interest.

**Data availability.** Data underlying the results presented in this paper are not publicly available at this time but may be obtained from the authors upon reasonable request.